# Embodied Consciousness Theory

Jahan N. Schad, PhD; Retired LBNL (UCB) Scientist
376 Tharp Drive, Moraga, Ca 94556

Email: schadn5@berkeley.edu

Phone: (925) 376-4126

The mysterious phenomenon of consciousness, after having been the subject of philosophic attention for few millennia, has drawn much scientific curiosity in recent decades; and many brilliant minds of various areas of sciences are trying to throw some light on it. Present neuroscience knowledge allows envisioning the neural processes behind the physical aspects of consciousness, however what may be behind the experiences of it, which has no physical insignia, has remained a mystery despite the grounds gained in recent well received efforts: monumental endeavors concluding in the known theories such as global work space theory (GWT) [1] and integrated information theory (IIT) [2], still fall short of providing a solid theory for consciousness: the former "proposes a simple hypothesis concerning the neural basis of ''making a conscious mental effort," and the latter, assuming experience to be an intrinsic property of the brain, "formulates how it is transitioned, through certain information based neural activity, to its physical substrate." Present work puts forward a theory of consciousness, rooted in the simple and straightforward implications of the computational operation of the brain which is consistent with well known facts and recent findings, which indicates presence of motor cortex activity in relaying conscious experiences. Despite simplicity, the theory provides fundamental basis for physicalism [3], as well answers for some Meta level problems of consciousness.

The Concept

Consciousness, the perceptions/conceptions of the inner/outer environment, as well as the commensurate activities, is rendered by the computations (simulations) in the brain-- evoked by the inputs from our five senses (at least four are of tactile nature) -- should inferentially be considered the expressions of the outputs at the interfaces available to the brain; and certainly in case of the physical expressions, the interface is the body. The physical aspects of consciousness are promulgated by the efferent signals, which are known to be issued from various parts of the motor cortex, graphically depicted by the homunculus despite the inaccuracy in suggesting the concept of the specific modular brain functioning; other parts of the brain are normally engaged in rendition of most effects. However, some subjective aspects of it (the non-physical experiences) also promote the physical effects: such facial and bodily displays of "emotions" have been noted since ancient times, and the following quote from St. Augustine [4] vividly depicts it:

*"And that they meant this thing and no other was plain from the motion of their body, the natural language, as it were, of all nations, expressed by the countenance, glances of the eye, gestures of*

*the limbs, and tone of voice, indicating the affections of the mind, as it pursues, possesses, rejects, or shuns."*

And therefore this fact attests to the presence of the motor neuron signals in at least in some aspects of our "subjective experiences;" the hard part of consciousness phenomenon. To this point, the report "reproducibility" of some of the subjective experiences, which is a "physical activity," further confirms it. Also, there is some evidence of "motor activity along with perceptual and semantic processes," which is reported in the global work theory (GWT) [1], published in the proceeding of the national academy of sciences (PNAS). The direct evidence of the continued activity of the motor neurons is clearly reported in the following quotes from research in Motor Robotics [5]

*"Motor cortex is also engaged when actions are observed or imagined."*

*"Beyond its central role in movement generation, primary motor cortex (MI) also appears to be engaged in cognitive and sensory processes in the absence of overt movement"*

Engagement of sensory cortex in "observation" and motor cortex in "physical activity" is expected, however, presence of "motor neuron activity when observing", with no inclining of even imagining action, is unexpected because of the "efferent nature" of such download even in the face of no apparent physical activity. The above fact provide the context in which the disruptive concept of the embodied consciousness based on simple inference from ground level understanding of the computational brain, can be formalized: This falsifiable theory put forward here, claims that consciousness in all its totality, physical, reported and unreported subjective aspects, all are bound to be "downloads, in the realm of the computational operations (input-process-output), of brain" by means of neural signals (some proven motor) to the interfaces of the body; some in muscular- skeletal displays, some in vocalization and some inevitably in the likeness of thought signals. And thought signal are also very likely to be of the efferent nature, which is indicated by the bimodal (on-off) vocal nature of thoughts (some people do their thinking loud all the time). Some of the efferent signal expressions are undoubtedly filtered by the bandpass characteristics of the bodily interfaces, which have limited the development of syntactic expression for the quality of the Qualia; and much is inhibited to begin with, with options for intentional display, as in vocalization. The embodied consciousness theory (ECT) is predictive of permanent activity of motor cortex, in conjunction with other areas of the cortex; in waking hours and REM sleep, whether one gets note of it or not, since many activities occur autonomously.

Another major aspect of the mystery of the consciousness phenomenon, which is comfortably avoided in the context of "one is one's brain," or presumed to have been answered in the vernacular of the IIT and GWT, is, how "persona" the identity of the claimant of consciousness, has come about; a sort of "meta problem of consciousness," perhaps similar to the other meta problem regarding the validity of the very question of consciousness itself that philosophers have raised [e.g., 6]; the latter a likely by-pass of the main question which may possibly resolve the mystery, or do away with it all, in approaches such as illusionism [7]. However, the ECT suggests a ground level explanation for the both Meta aspects in the context of brain's underlying dynamics that has led to the very development of the "referential" language syntaxes [8], serving as a basis for the development of the persona identity, and perhaps as well for the d3eveopmet of the spoken language. To this end, we need to start with the fact that whatever detailed information (afferent signals) related the characteristics of "things out there (not necessarily in themselves)--" which are parts of existence -- that our senses (vision inclusive) relay to the brain, has to result in the specific outputs rudely defining the "things" from which senses signals are picked up. It must be obvious that each "thing" (characteristic) depiction, the results of the brain computations, will have its nature stamped in the variances of the outputs. Parts of these characteristics, downloaded through bodily interfaces end up in specific vocalization (in case of absence of the vocal box, bodily expression or chemical outputs do the function), which gradually gets refined through the evolutionary processes, becoming the referral identity of "things"; the specific "referential" calls. And it is not hard to speculate the referential language much ties in with survival drive of beings. Also, the vocalizing body, being always in internal physiologic commotion (one way or the other), would have its own specific generally fixed download from the brain, which would always appear distinctly separately, or as a "preamble" to whatever else gets downloaded in the vocal box; and this permanent signature, as preamble or otherwise, becomes the identity of the being, distinguishing one from another, "ones Identity (the Is); somewhat like the "MAC" address of data processing devices which identifies them!

Conclusion

Our computational brains, which process the data received through sensation by our five interfaces with the environment,-- deploying learned and inherited neural patterns, and/or trial and error adjustments of synaptic functions-- also necessitates presence of interfaces for the expressions of the outputs. As known, some physical, as wells as some "emotional" aspects of our consciousness are broadcasted through our body interfaces; muscular-skeletal for the former and mostly facial expression for the latter, which are indicative of motor signals. The acts of "conscious experience reporting", and "loud thinking (voluntary or otherwise)", evince motor cortex involvement promulgating the act. These facts, supported by the recent research indicative of motor cortex activity during simple "observation" where no action, even in the realm of imagination, is intended or involved, provides the basis for theorizing that the output of brain computation are motor signals, which engages the body, though mostly vocal system, for its expressions. In this context, thoughts are low energy signal or perhaps of certain frequency range that is outside of bandpass of the vocal box, and therefore remains muffled. This testable theory claims that consciousness is embodied.